\documentclass{article}
\usepackage{pdfpages}
\usepackage[]{hyperref}
\usepackage{acra}
\usepackage{graphicx}

\title{Rage Music Classification \& Analysis using K-Nearest Neighbour, Random Forest, Support Vector Machine, Convolutional Neural Networks, and Gradient Boosting}
\author{Akul Kumar}

\begin{document}

\maketitle

\begin{abstract}
In this paper, we classify rage music (a subgenre of rap well-known for disagreements on whether
a particular song is part of the genre) with an extensive feature set through algorithms including Random
Forest, Support Vector Machine, K-nearest Neighbour, Gradient Boosting, and Convolutional
Neural Networks. We compare methods of classification in the application of audio
analysis with machine learning and identify optimal models. We then analyze the significant
audio features present in and most effective in categorizing rage music, while also identifying key audio features as well as broader separating sonic variations and trends.\end{abstract}

\section{Introduction}

The classification of music into distinct genres and emotional categories has long been a subject of interest in both musicology and machine learning.

This research focuses on a particularly intriguing genre: rage music. Characterized by real world debates on if certain songs are rage songs or just part of the wider trap genre, rage music presents both challenges as well as opportunities in machine learning. 

Our study aims to develop and compare various machine learning models for the task of classifying rage music, with a particular emphasis on the sonic features we apply. We explore a range of classification techniques, including Random Forests, Support Vector Machines (SVM), K-Nearest Neighbors (KNN), Convolutional Neural Networks (CNNs), and Gradient Boosting.

The core of our analysis lies in the extraction and interpretation of audio features that capture the essence of the genre. These features span a wide range of musical characteristics, from tempo and beat strength to spectral properties and harmonic content. By applying advanced machine learning techniques to these features, we seek to uncover the underlying patterns that distinguish rage music from other genres. We focus on identifying key characteristics and the most important parameters of successful/popular rage songs. 

\section{Methodology}
For all learning methods, the audio dataset was preprocessed using the librosa library, a popular tool for music and audio analysis. Features included Tempo (BPM), Beat strength, Onset rate, RMS energy, Spectral centroid, Spectral rolloff, Spectral flatness (as a measure of saw wave intensity), Zero-crossing rate, Mel-Frequency Cepstral Coefficients (MFCCs), 13 MFCC coefficients, Chroma mean, Chroma standard deviation, and Pitch features, amongst others. These features were chosen to capture a wide range of acoustic properties relevant to the classification task, especially due to the distinct uses of saw-waves in rage music. For a complete feature set, see the supplementary materials section of this paper. 

The dataset consisted of 1236 distinct audio files, with an overall duration of 4326 minutes. The non-rage class consisted of genres both distant to and sonically similar to rage music: there was a larger presence of similar genres such as dark trap, punk rock, punk rap, and heavy metal, while there was a smaller presence of distant genres such as pop, folk, and indie. Songs were selected on the basis of community agreement with status; for example, `Vamp Anthem' by Playboi Carti was selected due to its listing as such on informational websites and spotify status. Conversely, `Neon Guts' by Lil Uzi Vert was not selected due to its `trap' status on spotify and lack of rage listings on informational websites. The relatively small sample size is due to the lack of available songs; rage only recently emerged as a subgenre of rap in 2020 and there are few albums which are uncontested as rage music. For a complete list of audio files and genre-wide distribution, please see the supplementary materials section of this paper. 

The performance of each model was evaluated using standard metrics including accuracy, precision, recall, and F1-score. Additionally, Receiver Operating Characteristic (ROC) curves and Area Under the Curve (AUC) were computed to assess the models' discriminative capabilities. Confusion matrices were generated to visualize the classification results and identify patterns of misclassification. This multi-model approach allows for a comprehensive comparison of different machine learning techniques in audio classification, providing insights into the strengths and limitations of each method in the context of the detection of rage music. 

A diverse range of music files were used for both the rage music and non-rage music classifications. An extensive list of genres were compared against rage music, with a larger presence of sonically similar genres such as heavy metal, death metal, and dark trap due to the difficulty in separating boundary cases. There was a far smaller selection of distant genres such as folk and indie. For a comprehensive list of songs and artists used in both categories, please see the supplemental materials section of this paper. 

We employed a comprehensive set of evaluation metrics to assess and compare the performance of our models, including accuracy, precision, recall, F-1 score, Area Under the Receiver Operating Characteristic Curve, Cohen's Kappa, Matthews Correlation Coefficient, and Log Loss.
\subsection{Hyperparameter Tuning}
We employed a rigorous approach to model training and hyperparameter tuning: The dataset was split into training (80\%) and testing (20\%) sets using stratified sampling to maintain class balance. We used 5-fold cross-validation during the training process to ensure robust performance estimation.
For the Random Forest model, we performed a grid search over the following hyperparameters:
\newline
\begin{table}[ht]
\centering
\begin{tabular}{|l|c|}
\hline
\textbf{Hyperparameter} & \textbf{Values Tested} \\ \hline
n-estimators & [50, 100, 200] \\ \hline
max-depth & [None, 10, 20] \\ \hline
min-samples-split & [2, 5,, 10] \\ \hline
min-samples-leaf & [1, 2, 4]  \\ \hline
\end{tabular}
\end{table}
The optimal hyperparameters were selected based on the mean cross-validation score, and are provided below: 
\begin{table}[ht]
\centering
\begin{tabular}{|l|c|}
\hline
\textbf{Hyperparameter} & \textbf{Value} \\ \hline
n-estimators & 200 \\ \hline
max-depth & None \\ \hline
min-samples-split & 5 \\ \hline
min-samples-leaf & 2  \\ \hline
\end{tabular}
\end{table}

For all other learning models, the same approach (a grid search) was followed and optimal hyperparameters were selected on the basis of the mean cross validation scores. 
\subsection{Feature Importance \& Model Interpretability}
To gain insights into the most influential features for rage music classification, we employed several techniques. We extracted feature importances from the Random Forest model, which are computed based on the mean decrease in impurity across all trees. We also calculated SHAP values to understand the impact of each feature on model output across the entire dataset. We then generated partial dependence plots to visualize the marginal effect of selected features on the predicted outcome. 
 
We conducted statistical tests to further validate our findings, performing independent t-tests for each feature to determine if there were statistically significant differences between rage and non-rage music samples. Also, we computed a correlation matrix to understand the relationships between different audio features.

By combining these methodological approaches, we aimed to provide a comprehensive and rigorous analysis of rage music classification, grounded in both machine learning techniques and musical understanding.
\section{Results \& Findings}
\subsection{Model Performance \& Comparison}
K-Nearest Neighbour was the best model by all metrics (with an overall 0.9431 accuracy), followed by Random Forest (figure 1). 
\begin{figure} [ht]
    \centering
    \includegraphics[width=0.8\linewidth]{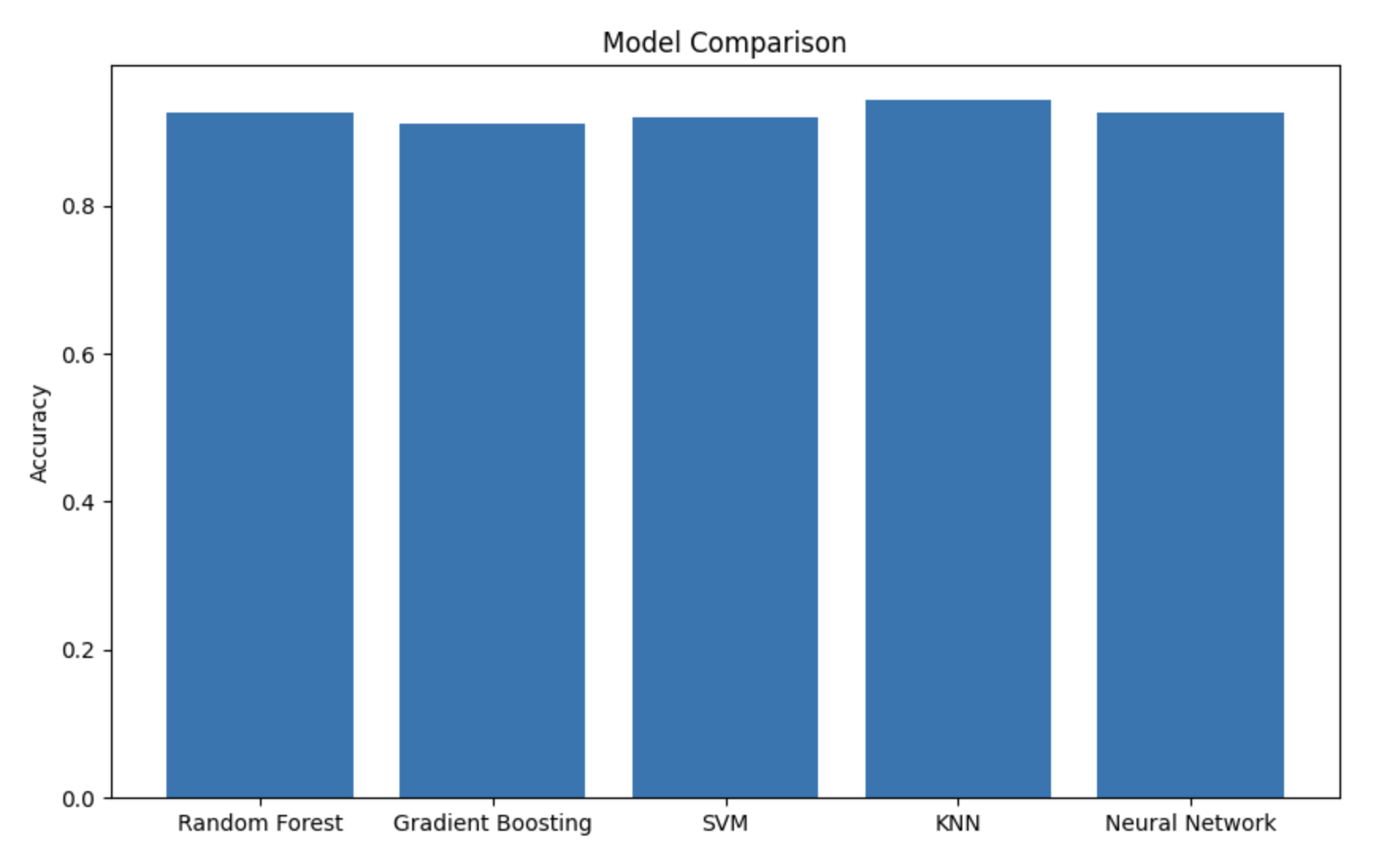}
    \caption{Accuracy across Different Models}
    \label{fig:enter-label}
\end{figure}

Generally speaking, non-linear classifiers outperformed linear models such as Support Vector Machine classification (SVM). This is supported by the PCA vs tSNE visualization (figure 2).  
\begin{figure} [ht]
    \centering
    \includegraphics[width=1\linewidth]{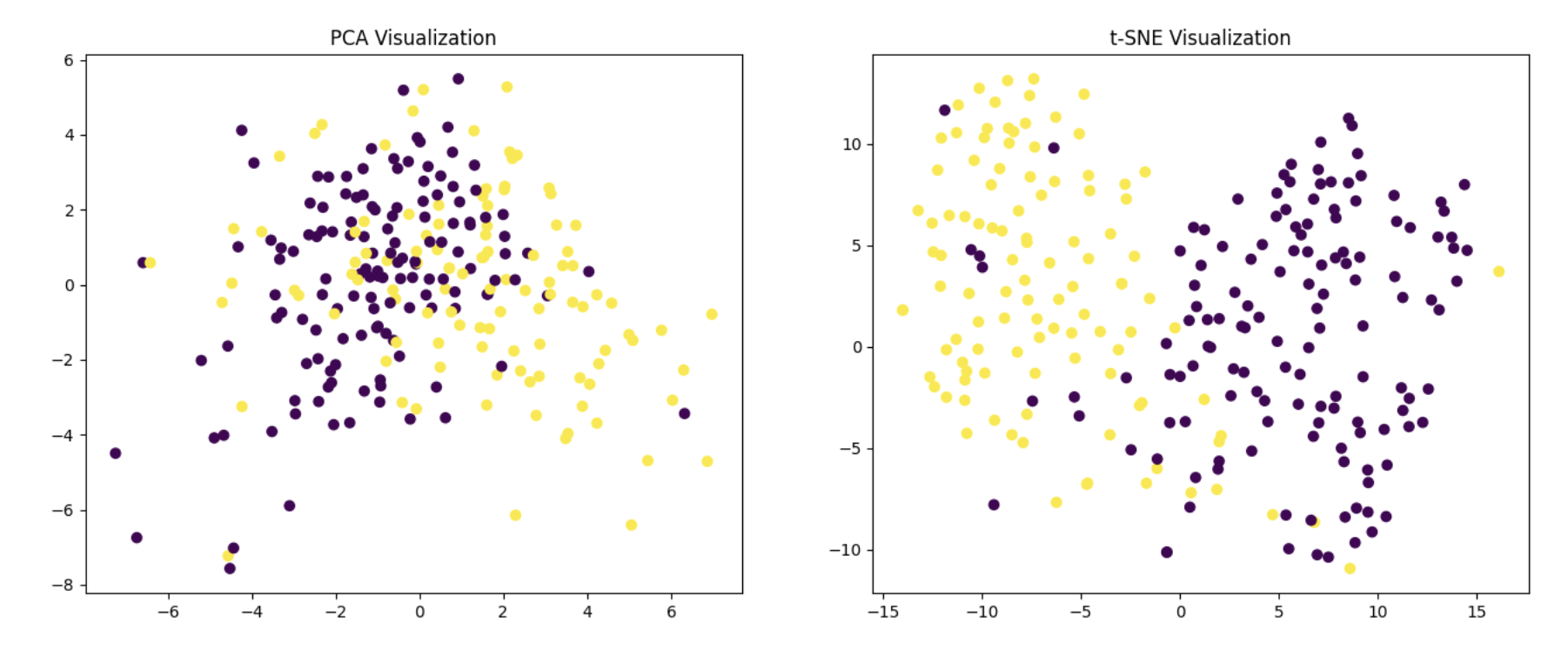}
    \caption{PCA and tSNE Visualizations}
    \label{fig:enter-label}
\end{figure}

The PCA vs tSNE visualization makes the intriguing classification problem clear; there is significant overlap between yellow (rage) and purple (non-rage) points. This implies linear separability in this 2D projection is low, suggesting potential challenges for linear classifiers. 

Overall distribution shows a clearer separation between classes compared to PCA. This suggests non-linear relationships in the original feature space are crucial for discrimination. PCA's linear projection shows high class overlap, while t-SNE's non-linear mapping reveals more separable structures.

The multiple clusters in t-SNE suggest that both rage and non-rage categories might benefit from hierarchical or sub-category classification approaches, and that linear classifiers such as logistic regression and linear SVM might struggle (which supports our observations regarding SVM performance). The effectiveness of t-SNE in revealing structure suggests that other manifold learning techniques (e.g. UMAP and Isomap) are likely to increase overall performance when used as a preprocessing step for classification.
\begin{figure} [ht]
    \centering
    \includegraphics[width=0.9\linewidth]{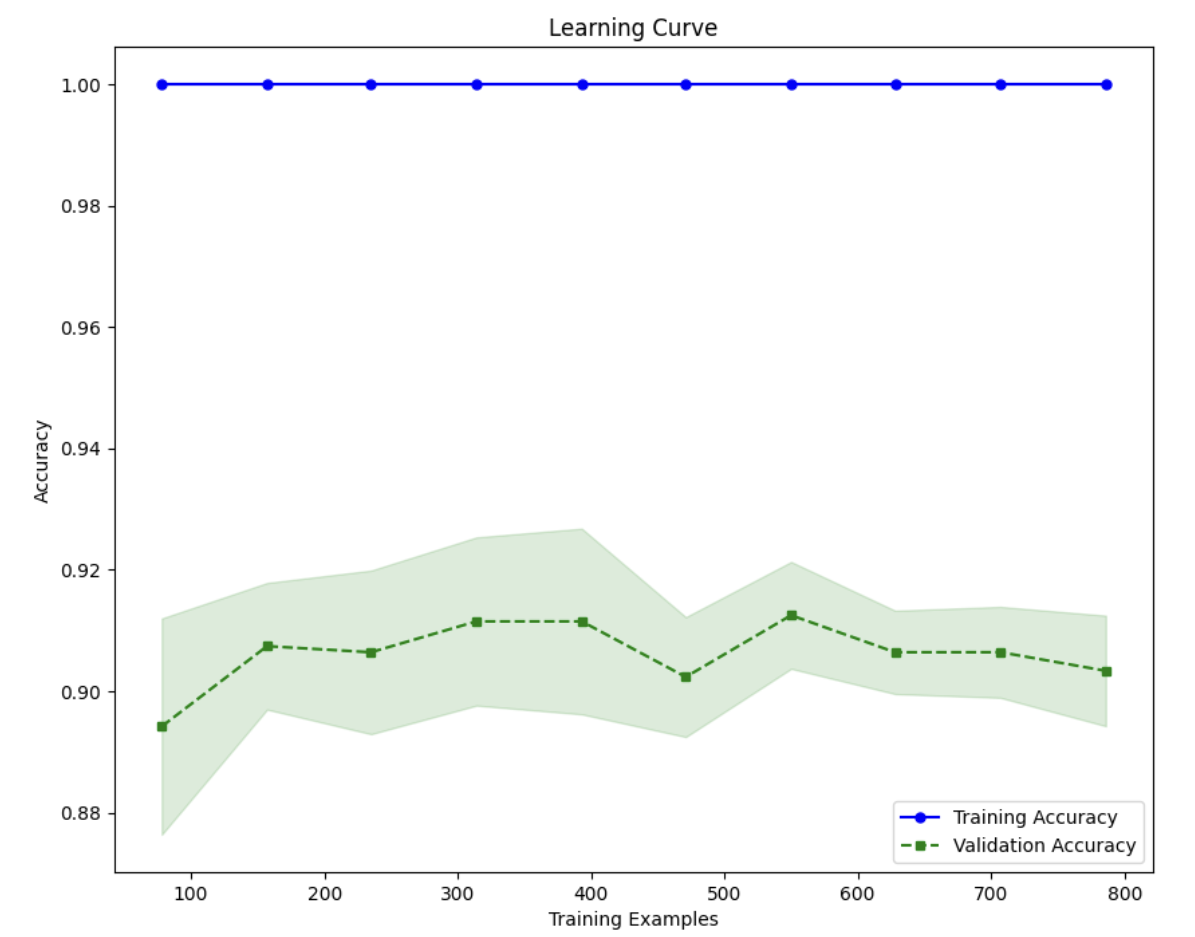}
    \caption{The Learning Curve for our RF Model}
    \label{fig:enter-label}
\end{figure}

The validation accuracy is 89.5\% at 100 training examples, and peaks at 91.2\% from 300-400 training examples
It then settles to approximately 90.5\% for larger training sets. There is a persistent gap of about 9-10\% between training and validation accuracy - this 
gap doesn't narrow significantly as training data increases. 

The model achieves near-optimal performance with relatively few examples (300-400). The perfect training accuracy coupled with the persistent gap to validation accuracy suggests some overfitting, and the flat validation accuracy after 400 examples suggests that model fusion is needed to drive a significant change in future performance. 

It is evident from the calibration curve (figure 4) that the model has a very sharp decision boundary around 0.4 predicted probability. This suggests that there are distinct, identifiable features that separate rage from non-rage music, correlating with greater MFCC 3 and MFCC 10 values (see section 3.2). The jumps in the calibration curve correspond to critical thresholds for key features; for example, specific harmonic to percussive ratio values. The sharp jump indicates that the model’s probability outputs in the middle range (0.3-0.5) are not well-calibrated. For applications requiring
accurate probability estimates (rather than just binary classification), the model would benefit from
calibration techniques like Platt scaling or isotonic regression.
\begin{figure}
    \centering
    \includegraphics[width=1\linewidth]{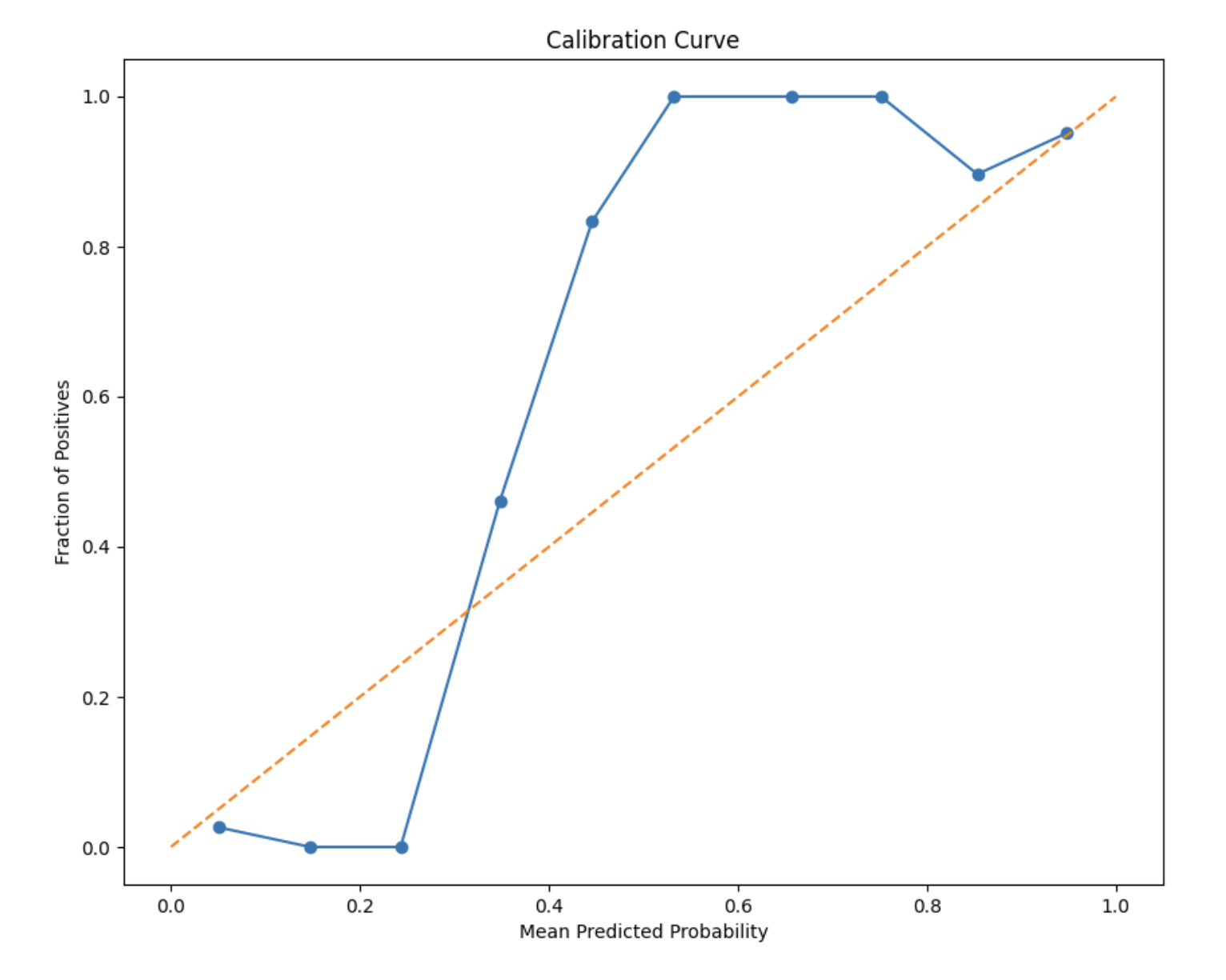}
    \caption{The Calibration Curve for our Model}
    \label{fig:enter-label}
\end{figure}

\subsection{Feature Importance \& Analysis}
The most significant features remained largely constant amongst all models (figure 5), with the ordered 5 most important SHAP (SHapely Additive exPlanations) features being: song length, harmonic ratio, percussive ratio, chroma mean, and Mel Frequency Cepstral Coefficient 3 (MFCC3).
\begin{figure} [ht] 
    \centering
    \includegraphics[width=1\linewidth]{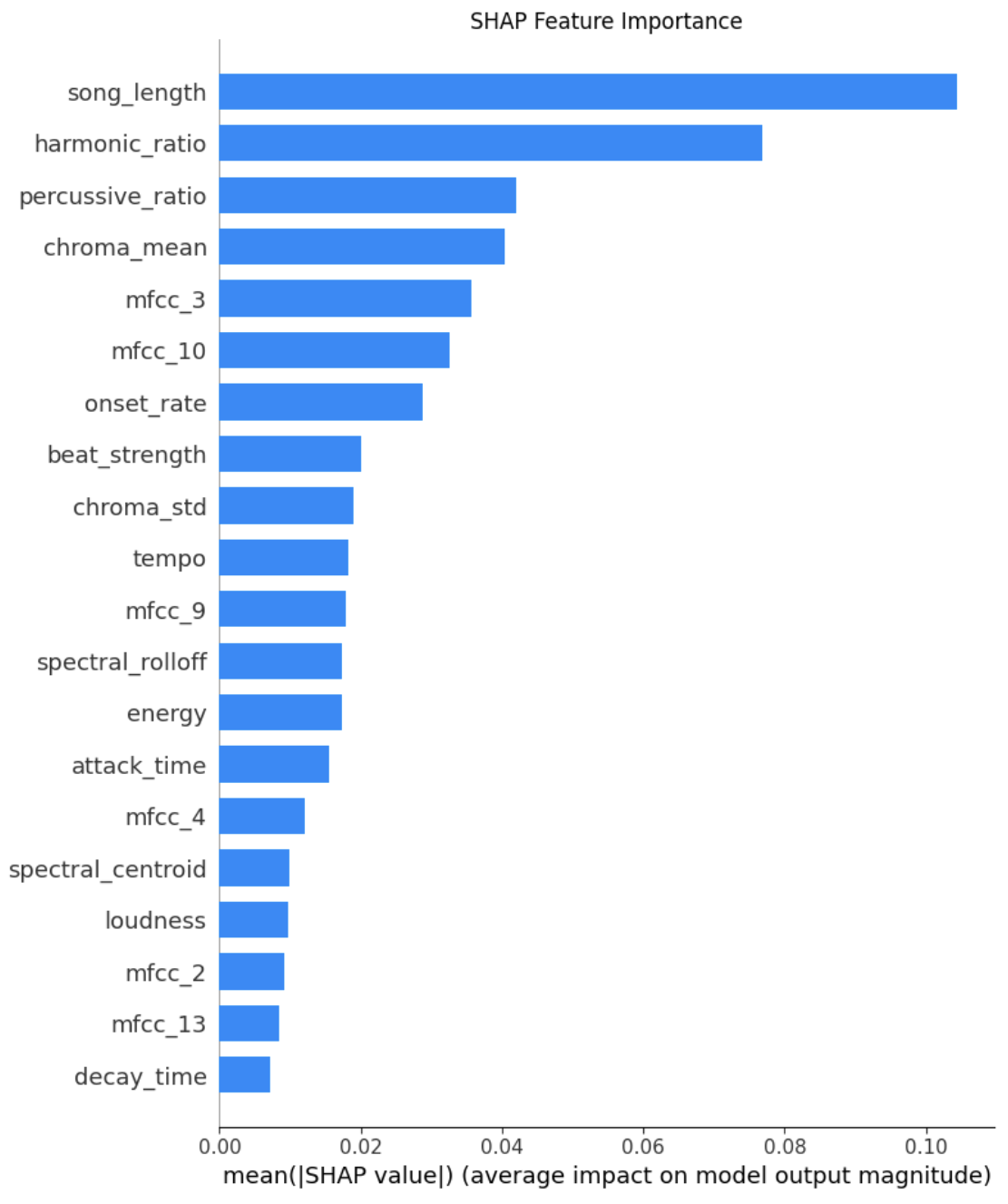}
    \caption{SHAP Feature Importance}
    \label{fig:enter-label}
\end{figure}

Interestingly, song length is the most important feature according to SHAP values. As the top feature, song length reveals distinctive structural conventions in rage music. Our data indicates a bimodal distribution, with short, intense tracks (1.5-2.5 minutes); and extended compositions (4-6 minutes). 

The presence of spectral rolloff and spectral centroid in the top features indicates that the distribution of frequencies is a key identifier. This relates to the level of distortion or the presence of specific synthesizer tones characteristic of the genre. Rage music’s characteristic high vocal inflections often cause the spectral centroid to (on average) exceed 2kHz. On the other end, low pitch inflections such as those in Playboi Carti’s 'On That Time' are characterized by stronger lower formants (500-1000 Hz).

Suprisingly, tempo is only moderately important. This suggests that while rage music might have a characteristic tempo range/threshold (see the discussion on partial dependence curves), it's not as defining as other features. This, along with the low importance of mean pitch and the standard deviation of pitch indicates that rage music is more defined by its timbral and rhythmic qualities than by specific pitch patterns of melodic structures. 

Chroma features (mean and standard deviation) are among the top predictors for rage music classification. This suggests specific harmonic patterns characteristic of the genre - the prominence of chroma mean indicates a preference for certain pitch classes. The importance of chroma standard deviation suggests that rage music, while often perceived as harmonically simple, actually exhibits significant harmonic variation This could manifest as rapid chord progressions, frequent key changes or modulations, and the use of chroma passages or dissonant intervals. The combination of chroma features points to a  prevalence of Phrygian and Locrian modes, known for their dissonant and tense qualities. These modes, particularly the Phrygian dominant scale, are often associated with aggressive and intense musical expressions, which rage music embodies. 

The high important of MFCC 3 (Mel-frequency cepstral coefficient 3) suggests a predominance of distorted vocal techniques, with rapid alternations between clean and distorted vocals. Multiple MFCC features (3, 10, 9, 4, 2, 13) appear in the top 20 SHAP features. This highlights the importance of timbral characteristics in identifying rage music. Specific MFCC coefficients capture the unique textures of distorted 808s, timbral qualities of the aggressive vocal delivery, and 
characteristics of the synthesizer sounds frequently used in the genre.

The harmonic ratio and MFCC features also indicate distinctive sonic features, including prominent mid-range frequencies (1-3 kHz) for cutting through dense mixes. The percussive ratio  coupled with the onset rate indicates greater consistency and power, with the prominent use of double bass drum patterns (150-200 BPM for sixteenth notes).  The eminent presence of significant high-mid energy frequencies (3-5 kHz) are caused by snare drums, used for cut and aggression. 

The high importance of both harmonic ratio (2nd) and percussive ratio (3rd) in the SHAP feature importance suggests that rage music has a distinctive balance between tonal and rhythmic elements. This could reflect the heavy use of distorted bass and synthesizers (harmonic content); prominent, aggressive drum patterns (percussive content); and specific mix balances which set rage apart from related genres such as trap.

The feature importance is also indicative of sidechain compression (particularly on bass) keyed to the kick drum, also signalling minimal reverb usage and layered distortion effects (both analog-modeled and digital distortions). 
\subsection{Partial Dependence Plot Analysis}
\begin{figure} [ht]
    \centering
    \includegraphics[width=0.9\linewidth]{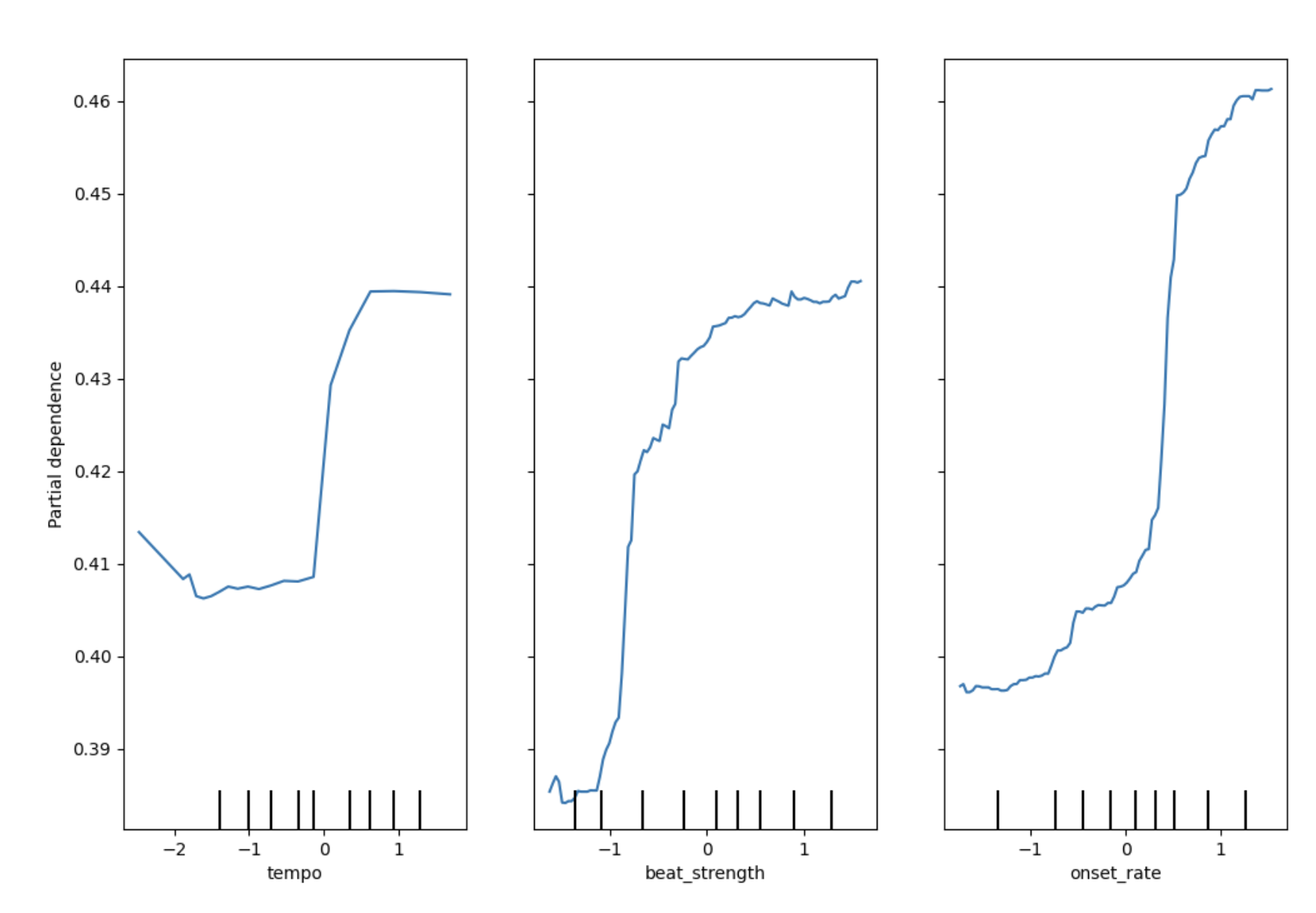}
    \caption{Partial Dependence Plots for Tempo, Beat Strength, and Onset Rate}
    \label{fig:enter-label}
\end{figure}

The partial dependence plot for tempo (figure 6) shows a distinct threshold effect. There is a sharp increase in the
model’s prediction of rage music at a specific tempo range, approximately centered around the 0 point on
the normalized scale (around 150-160 BPM). This tempo range is faster than general trap music (usually
around 130-140 BPM), with which it is often confused.

The tempo feature shows a distinct step function with three main regions:
\newline
a) Below -0.5: Low probability of rage classification at 0.407.
\newline
b) Between -0.5 and 0: Sharp increase in probability, rising to 0.438.
\newline
c) Above 0: Plateau at the higher probability of 0.439.

This step function suggests a critical tempo threshold for rage music. Since the normalization is
centered around the mean tempo of the dataset, we can infer that rage music has a characteristic
minimum tempo (151 BPM). Once this tempo threshold is crossed, the likelihood of a track being classified
as rage music increases significantly (by about 7.8\%). Beyond this threshold, further increases in tempo
don’t substantially affect the classification.
This aligns with rage music’s reputation for high energy and fast-paced beats. It suggests that while tempo
is important for defining the genre, it’s more of a prerequisite than a sliding scale - tracks need to be ”fast
enough” to be considered rage, but being extremely fast does not necessarily make a track more ”rage-like”.
  
Both beat strength and onset rate also show positive relationships with partial dependence. This positive relationship of partial dependency with onset rates signals that rage music is characterized by a high density of musical events or rapid rhythmic changes. The partial dependence plot for onset rate shows a strong positive correlation with the likelihood of classifying a track as rage music, particularly at higher values. The plot demonstrates a steep increase in
partial dependence from approximately 0.40 to 0.46 as the onset rate increases. 
This suggests that rage music is characterized by a high density of note onsets, which could manifest as rapid-fire lyrical delivery (consistent with the genre’s association with fast, aggressive rap styles); complex, busy instrumental
patterns - possibly including rapid hi-hat rolls, quick synth arpeggios, or densely layered production
elements; or frequent rhythmic changes or breaks.

The beat strength plot shows a more gradual, mostly monotonic increase:
\newline
a) Sharp rise from -1 to about -0.75. \newline
b) Relatively steady increase from -0.75 to about 0.5.\newline
c) Steeper increase from 0.5 to 1. 

This suggests that stronger beats consistently increase the likelihood of a track being classified as rage music. The steeper increase at higher beat strengths implies that very strong, prominent beats are particularly characteristic of the genre. The overall change in probability (from about 0.385 to 0.44, or about 14.3\% increase) is larger than for tempo, indicating that beat strength is a more influential feature in defining rage music.

The onset rate plot shows the most dramatic and consistent increase:
\newline
a) Steady, almost linear increase from -1 to about 0.5.\newline
b) Steeper increase from 0.5 to 1. 

The probability of rage classification
increases from about 0.395 to 0.465,
a substantial 17.\% rise. This is the largest increase among the three features, highlighting the critical
importance of onset rate in defining rage music.
The near-linear relationship suggests that, unlike tempo, onset rate contributes to the energy of a track
across its entire range. Higher onset rates consistently make a track more likely to be classified into the rage genre. 

Onset rate has the widest range of influence on classification probability (0.07 change), followed by
beat strength (0.055 change), and then tempo (0.032 change). Onset rate is the most consistently
influential, suggesting that the density of musical events (be it in the percussion, synths, or vocals) is a key
defining feature of rage music.
\section{Conclusion}
This paper analyzes the effectiveness and performance of different machine learning algorithms in the problem of classifying rage music, showing that the K-Nearest Neighbour algorithm  is optimal for this application. We also observe and conclude that non-linear classifiers (such as kernel SVM) may struggle with music genre classification, and similar applications likely benefit from hierarchical or sub-category classification. 

Furthermore, it finds that the distribution of song length amongst rage songs is bimodal and that song length has the greatest defining impact on the genre when compared to other sonic and temporal features. We also observe that the genre is characterized by certain pitch classes (due to the observed importance of chroma mean) and frequency distributions. The observed importance of chroma standard deviation suggests that rage music is not as harmonically simple as it is perceived and is harmonically complex. Overall, the importance of chroma features suggests that rage music is more dependent on its rhythmic/timbral qualities as compared to specific pitch patterns or melodic structures. 

The partial dependence plots created reveal that certain features such as beat strength are stronger indicators than tempo (inspite of the genre's signature high BPM) and that selected features including tempo exhibit a threshold effect. Our study also observes a high onset rate, indicating the high density of musical events in the genre.

\textbf{Limitations.} Despite the robustness shown by the performance of our models, for applications requiring accurate probability estimates (rather than just binary classification), the model would benefit from
calibration techniques like Platt scaling or isotonic regression. Also, the effectiveness of t-SNE in revealing structure suggests that other manifold learning techniques (e.g. UMAP and Isomap) are likely to increase overall performance when used as a preprocessing step for classification. 
\section{Supplementary Materials}
For a complete compendium of extracted data, a list of audio files used in the initial dataset, and the original feature set, please consult \href{https://drive.google.com/drive/folders/14qKIwAXog9etLcc0oFnDKhgWvmADE4Ji?usp=sharing}{\underline{this}} link.


\begin{thebibliography}{99}

\bibitem{duda1996pattern}
Duda, R. O., \& Hart, P. E. (1996). Pattern recognition and scene analysis. \textit{Journal of the Brazilian Computer Society}, \textit{2}(2), 50-51. Springer. \url{https://journal-bcs.springeropen.com/articles/10.1007/BF03192561}

\bibitem{tindale2012machine}
Tindale, A., Wanderley, M. M., \& Malloch, J. (2012). Machine learning approaches for music information retrieval. In \textit{Proceedings of the 2012 International Conference on New Interfaces for Musical Expression (NIME 2012)} (pp. 64-69). ResearchGate. \url{https://www.researchgate.net/publication/221787719_Machine_Learning_Approaches_for_Music_Information_Retrieval}

\bibitem{wu2021multi}
Wu, Y., Zhang, Y., \& Yang, Q. (2021). Multi-task learning for music processing. \textit{arXiv preprint arXiv:2107.09208}. \url{https://arxiv.org/pdf/2107.09208}

\bibitem{peeters2007audio}
Peeters, G., \& Flocon-Cholet, C. (2007). Audio onset detection using machine learning techniques: The effect and applicability of key and tempo information. In \textit{Proceedings of the 8th International Conference on Music Information Retrieval (ISMIR 2007)} (pp. 535-536). Academia. \url{https://www.academia.edu/31215219/Audio_onset_detection_using_machine_learning_techniques_the_effect_and_applicability_of_key_and_tempo_information}

\bibitem{allen2016word}
Allen, H., \& Hampapuram, S. (2016). Word embeddings for music recommendation. \textit{CS224D: Deep Learning for Natural Language Processing, Stanford University}. \url{https://cs224d.stanford.edu/reports/allenh.pdf}

\bibitem{liang2022deep}
Liang, D., Hawthorne, C., Stober, S., \& Sapp, C. (2022). Deep learning for musical form recognition and analysis. ResearchGate. \url{https://www.researchgate.net/publication/361023801_Deep_Learning_for_Musical_Form_Recognition_and_Analysis}

\end{thebibliography}
\end{document}